\documentclass[notitlepage,letterpaper,12pt]{article}
\usepackage[pdftex]{graphicx}
\usepackage{abstract}
\usepackage{hyperref}

\newcommand\myurl[2]{\url{#1}}
\newcommand\email[1]{\href{mailto:#1}{\nolinkurl{#1}}}

\title{The Need for an R\&D and Upgrade Program for CMS Software and Computing}

\author{Peter Elmer\\
Princeton University\\
\email{Peter.Elmer@cern.ch} \and
Salvatore Rappoccio \\
SUNY - Buffalo \\
\email{srrappoc@buffalo.edu} \and
Kevin Stenson \\
University of Colorado - Boulder \\
\email{Kevin.Stenson@colorado.edu} \and
Peter Wittich \\
Cornell University \\
\email{wittich@cornell.edu}}
\date{\today}

\begin{document}

\maketitle
\thispagestyle{empty}

\begin{abstract}
  Over the next ten years, the physics reach of the Large Hadron
  Collider (LHC) at the European Organization for Nuclear Research
  (CERN) will be greatly extended through increases in the
  instantaneous luminosity of the accelerator and large increases in
  the amount of collected data. Due to changes in the way Moore's Law
  computing performance gains have been realized in the past decade,
  an aggressive program of R\&D is needed to ensure that the computing
  capability of CMS will be up to the task of collecting and analyzing
  this data. This white paper is intended to motivate such a program.

  To achieve the Compact Muon Solenoid collaboration's physics goals,
  a computing system of greatly increased performance will be required
  to process, simulate and analyze the resulting increase in data
  volume.  
  In the past, CMS computing has been able to rely on industry to
  deliver exponential increases in performance per unit cost over
  time, as famously described by Moore's Law.  

  Realizing these exponential gains in processor performance per unit
  cost will be much more difficult in the future than over the past
  few decades. In recent years, technology limitations, in particular
  regarding power consumption, have triggered profound changes in the
  evolution of computing processor technology. In the past software
  could be run unchanged on successive processor generations and
  achieve Moore's Law-like performance gains. This behavior has
  allowed software designs based on simple, sequential programming
  models to scale easily through enormous increases in
  performance. The era of scaling for such sequential applications is
  now over. The limitations on power consumption are leading to a new
  era in which scalability will need to be achieved via significantly
  more application parallelism and the exploitation of specialized
  floating point capabilities. Achieving these huge potential
  increases will transform completely the processor landscape and
  software design. Failure to adapt will imply an end to the
  exponential cost reductions for computing which have been
  fundamental to enabling the progress of science in general and
  specifically to the discovery program at the LHC and will be
  required to maintain current capabilities in the era of flat budgets
  and increased data complexity.

  Thus, in order to guarantee the success of our scientific program,
  a dedicated R\&D and upgrade program for software
  and computing is needed in parallel to the planned LHC and CMS
  detector upgrades over the next decade. A broad and balanced mix of
  effort on a number of elements will be required, including general
  investigations into newer processor architectures and programming
  models, the simulation, pattern recognition algorithms in the
  experiment trigger and reconstruction, tools and systems and
  analysis techniques. Many aspects of the areas to investigate are
  not unique to CMS, nor to HEP, but as always the needs of our
  scientific research program compel us to work at the leading edge of
  progress in computing technology. As deviations from Moore's Law
  cost scaling are already becoming visible, we expect that the
  efforts will result in concrete upgrades already in the next few
  years, however given the fundamental nature of the technology
  changes, these must be seen as steps along an R\&D path in the
  longer term eventually aimed at efficient scalability of our
  applications through order of magnitude increases in processor
  power.

\end{abstract}
\newpage

\section{Introduction}
This white paper describes the elements of an R\&D and upgrade
effort we believe will be required to meet future computing needs
for the Compact Muon Solenoid (CMS) experiment~\cite{CMSDET} at the
European Organization for Nuclear Research (CERN) Large Hadron
Collider (LHC)~\cite{LHCPAPER}. In the next ten years, increases
in instantaneous luminosity are expected that will greatly strain
the available computing resources. Simultaneously,  profound
changes in the evolution of computing processor technology imply
that the current software, without significant changes, is unlikely
to deliver sufficient performance and scalability on newer computing
hardware. A dedicated software and computing R\&D and upgrade effort,
in parallel to the CMS detector and LHC upgrades, will be required
to insure that we continue to benefit from the exponential gains
in performance/cost (Moore's Law) seen over the past decades and
that we are thus able to capitalize on and realize the full potential of
the investment in the LHC.

\section{Physics Motivation and LHC future plans}

The LHC is the largest scientific instrument ever constructed. With a
circumference of 17 miles, the apparatus straddles the border between
France and Switzerland.  At the LHC, we are recreating in the
laboratory conditions that have not existed since shortly after the
Big Bang by smashing together protons at the highest energy ever
achieved in the laboratory.  The energies we are exploring are
equivalent to the temperature of the universe one ten billionth of a
second after its creation.  The goal of the LHC research program is to
study the basic building blocks of matter and answer such questions
as: Is our universe super-symmetric? Where did all the antimatter go?
How do fundamental particles acquire mass?  The research focus
involves both studies of known standard model particles (such as the
top quark) and searches for new physics (such as supersymmetry or
extra space-time dimensions). In 2012, the LHC collaborations
announced the discovery of a Higgs-like Boson, marking the first in
what we expect will be a string of exciting discoveries.

CMS is one of the two general-purpose detectors at the LHC.
The central feature of the CMS apparatus is a superconducting solenoid
of 6~m internal diameter, providing a magnetic field of
3.8~T. Within the solenoid volume are a silicon
pixel and strip tracker, a lead tungstate crystal electromagnetic
calorimeter (ECAL), and a brass/scintillator hadron calorimeter
(HCAL). Muons are measured in gas-ionization detectors embedded in the
steel return yoke. Extensive forward calorimetry complements the
coverage provided by the barrel and endcap detectors.

The LHC beams are made to collide at a rate of 40~MHz; however, only a
few hundred Hz of these events can be saved to off-line storage for
later analysis. The decision whether to reject or keep an event must
be made within milliseconds of the collision, and requires a large and
sophisticated trigger computing farm to accomplish. Additionally, the
petabytes of data stored each year must be made available worldwide to
a collaboration of 3000 scientists for further refinement and analysis
via many offline analysis computing farms.  These farms are also used
to create simulated data samples at least ten times  larger than
the stored data to allow understanding of effects such as acceptance
and systematic uncertainties.  These tasks (trigger and offline
computing farms) must be provided at the lowest cost with the highest
performance possible.

A rough time progression of data-taking at the LHC follows. In
calendar year 2015, we will take data at full energy
($\sqrt{s}=13-14$~TeV) with a target instantaneous luminosity of
${\cal L}=1\times 10^{34}\mathrm{/cm^2/s}$. In 2018, there will be
another year-long shutdown. After this shutdown the machine will
deliver another factor of two increase in luminosity to ${\cal
L}=2\times 10^{34}\mathrm{/cm^2/s}$. The so-called High Luminosity
LHC (HL-LHC) phase will start in 2023, with another increase to ${
\cal L}=5\times 10^{34}\mathrm{/cm^2/s}$. With each increase in
luminosity, the number of additional soft collisions per bunch
crossing will increase, from the current level of 20-30 at ${\cal
L}=7\times 10^{33}\mathrm{/cm^2/s}$ to up to 140 during the
HL-LHC~\cite{HLLHC}, leading to an increase in processing time.
\begin{figure}[tbp]
\centering
\includegraphics[width=0.6\textwidth]{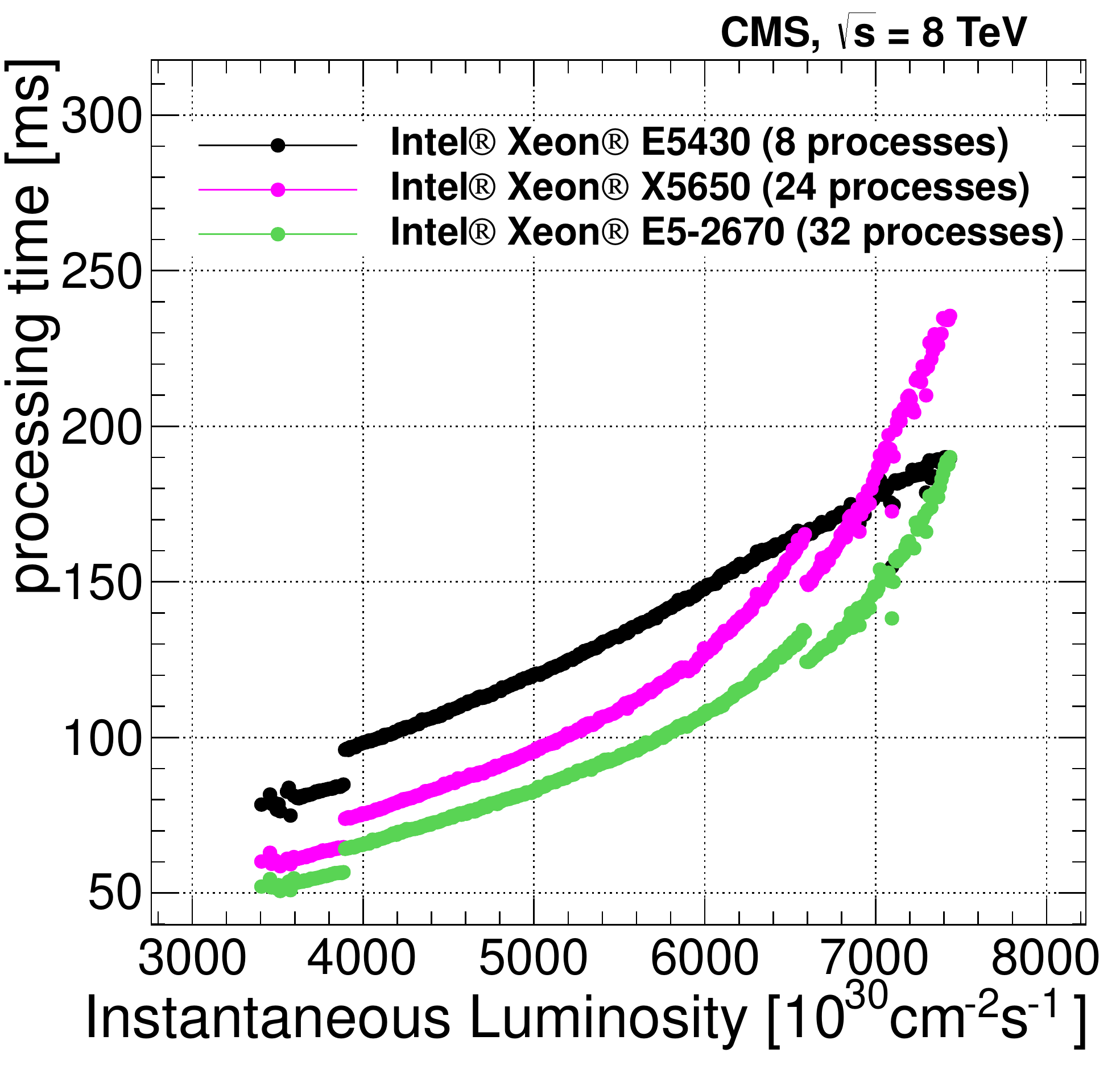}
\caption{HLT processing time per event as a function of
instantaneous luminosity on the three different machine types used
in the filter farm. The processing time increases more than linearly
with increasing instantaneous luminosity for the machines running 24
and 32 processes. } 
\label{fig:hltprocessing}
\end{figure}
As an example, Fig.~\ref{fig:hltprocessing} shows effects of increased
luminosity on the per-event processing time in the high-level trigger
(HLT). The more modern machines, which use Intel's HyperThreading
technology, show better performance overall but a faster-than-linear
increase in processing time with increased luminosity. We expect this
trend to continue.


\section{Overview of Industry Trends}

Recent years have seen a significant change in the
evolution of processor design relative to the previous decades~\cite{GAMEOVER}.
Previously one could expect to take a given code, and often the
same binary executable, and run it with greater computational
performance on newer generations of processors with roughly exponential
gains over time as described by Moore's Law.  A combination of increased
instruction level parallelism and (in particular) processor clock frequency 
increases insured that expectations of such gains could be met in generation 
after generation of processors. Over the past 10 years, however, 
processors have begun to hit scaling limits, largely driven by overall 
power consumption.

The first large change in commercial processor products as a
result of these limits was the introduction of ``multicore'' CPUs,
with more than one functional processor on a chip. At the same time
clock frequencies ceased to increase with each processor generation and 
indeed were often reduced relative to the peak. The result of this was
one could no longer expect that single, sequential applications would
run faster on newer processors. However in the first approximation,
the individual cores in the multicore CPUs appeared more or less
like the single standalone processors used previously. Most large
scientific applications (HPC/parallel or high throughput) run in
any case on clusters and the additional cores are often simply
scheduled as if they were additional nodes in the cluster. This
allows overall throughput to continue to scale even if that of a
single application does not. It has several disadvantages, though,
in that a number of things that would have been roughly constant
over subsequent purchasing generations in a given cluster (with
a more or less fixed number of rack slots, say) now grow with each
generation of machines in the computer center. This includes the
total memory required in each box, the number of open files and/or
database connections, increasing number of independent (and incoherent)
I/O streams, the number of jobs handled by batch schedulers,
etc.  The specifics vary from application to application, but
potential difficulties in continually scaling these system parameters
puts some pressure on applications to make code changes in response,
for example by introducing thread-level parallelism where it did
not previously exist.

There is moreover a more general expectation that the limit of power
consumption on future Moore's Law scaling will lead to more profound
changes going forward. In particular, the power hungry x86-64
``large'' cores of today will likely be replaced wholly or in part by
simpler and less power hungry ``small'' cores. These smaller cores
effectively dial back some of the complexity added, at the expense of
increased power, in the period when industry was still making single
core performance scale with Moore's Law.  The result is expected to be
ever greater numbers of these smaller cores, perhaps with specialized
functions like large vector units, and typically with smaller memory
caches than the ``large'' cores.  Exploiting these devices fully will
also push applications to make larger structural code changes to
introduce significantly more fine-grained parallelism.

Although it is very hard to predict precisely where the market will wind up
in the long run, we already see several concrete examples which give 
indications as to the kinds of things that we will see going forward:

\begin{itemize}
\item Intel's Many Integrated Core (MIC) architecture, combining many smaller cores with very-wide SIMD units. The first commercial products (Xeon Phi) are in the form of a coprocessor and aimed at the HPC market.
\item Systems implementing the forthcoming ARMv8 64bit architecture.  Here the significant use of the ARM processor in low-power or embedded systems (e.g. mobile devices) positions it well to enter a server market dominated by the power limitations described above, even if the route it followed to get there differs from that of Intel. Intel is also preparing its own low power server
variants, hopefully leading to a competitive market with price benefits for
buyers.
\item General Purpose Graphics Processing Unit (GPGPU or GPU), such as
the Tesla accelerators from NVIDIA.
\end{itemize}

Overall the market is likely to see significantly more heterogeneity in
products than in the past couple of decades. Effectively exploiting these newer
architectures will require changes in the software to exhibit
significantly more parallelism at all levels, much improved locality 
of access to data in memory and attention to maximize floating point
performance.  Most of the scientific software and algorithms
in use today in experiments like CMS was designed for the sequential processor 
model in use for many decades and require significant re-engineering to meet
these requirements.

If we fail to meet the challenge of adapting the software, the cost of computing
required for the luminosity upgrades of the LHC will not profit from
Moore's Law cost reductions as in the past.
Already the market trend studies of CERN IT~\cite{CERNIT} , for example, indicate
that they should decrease their expectations of overall throughput/cost gains
from $\sim$40\% per year to 20\% per year
for typical low-end servers with multicore CPUs which we use for high throughput
computing. This corresponds to the ``doubling time'' for performance/cost
roughly increasing from 1.5 years to 3 years. Only by embracing the newer 
architectures are we likely to have sufficient computing power for our 
scientific goals over the next ten years.

It is also important to recognize that if we obtain a full Moore's Law-like
gain by doubling of performance every 1.5 years, we are talking about 2 orders
of magnitude over the next 10 years and 3-4 orders of magnitude through
the HL-LHC era. The era of scaling for sequential applications appears to be over.
Achieving these huge potential increases will likely transform completely the
processor landscape and software design. Investigations to upgrade the software 
to the near/medium term processor products should be seen as steps along
an R\&D path in the longer term eventually aimed at efficient scalability of
our applications through order of magnitude increases in processor power.

We note also that the challenging technology evolution described here is
of course an important facet of the problems facing the Exascale Computing
initiative. While we do not face the truly massive parallelization requirements of
traditional supercomputer applications, the nature of our highly complex software
applications and the need for high throughput brings other types of challenges.

\section{Overview of CMS and HEP Software}

To illustrate the nature of the CMS software problem, we now give further
details as to the software stack we use and provide some examples of 
where we expect software upgrades and/or R\&D will be required.

As a starting point we first note that during 
2012, the CPU usage for offline computing (non-trigger) activities was very 
roughly divided up as:

\begin{itemize}
\item $\sim$40\% event simulation, mostly dominated by \textsc{geant4}~\cite{GEANT4} for
performance critical portions of the application
\item $\sim$20\% event reconstruction (for real data) and pile-up mixing/digitization/reconstruction (for Monte Carlo events), dominated by code written by CMS, specialized for the CMS detector, with an important contribution for I/O from ROOT~\cite{ROOT}
\item $\sim$40\% mixed user analysis applications
\end{itemize}
In addition there is the High Level Trigger (HLT) farm, whose CPU use
is also dominated by code written by and specialized for CMS. A
significant fraction of the HLT code base is in common with the
offline computing mentioned above. These numbers will however evolve
in 2015 and beyond, in particular for the reconstruction and the HLT,
which as described above we expect will become relatively more costly
with increasing pile-up events. We expect the cost of event simulation
will scale to first order with the total number of events. The data
analysis is a mix of activities, some of which scale as the number of
events and some which scale as their complexity.

The code base written by CMS personnel and distributed as a software
releases is about 3.5M source lines of code (SLOC) in C++ as measured
by the SLOCCount\cite{SLOCCOUNT} tool. Contributions to the code in
the software release have been made by approximately 1000 individuals
over the past 10 years, with up to 280 committing changes in any given
month. In addition there are at least several MSLOC of ``User''
analysis code, which is not centrally maintained in a single release,
but relevant for physics results.  Also relevant are a small number of
additional software libraries, in particular scientific codes provided
by groups outside CMS and typically used also by other
experiments. Important examples include \textsc{geant4} (1.2 MSLOC),
ROOT (1.7 MSLOC) and others. This is the scale of code required given
the complexity of the CMS detector and LHC environment, and the
challenge of identifying exceedingly rare events such as those which
led to the discovery of a Higgs-like boson in
2012~\cite{CMSHIGGS}. The fact that so many individuals have
contributed code is an important feature: a great deal of expertise in
both aspects of the detector and the physics of LHC is thus expressed
through the software.  The evolving nature of both the LHC luminosity
(and energy), as well as our ever improving understanding of the CMS
detector, also pushes the software to continually evolve.

The codes and the data management are typically set up to exploit the 
embarrassingly parallel processing possible at the level of events, which are
independent of one another. The resulting batch jobs do not need to
communicate among themselves: each job reads some number of
events, one after another, and outputs processed versions of those events.
The codes used by CMS (and experimental HEP in general) also tend to lack
clear numerical ``kernels'' where optimization efforts can be focused. 
Given these characteristics they are more properly classified as
``high throughput computing'' (HTC) rather than ``high performance computing'' (HPC). 
In terms of their detailed behavior on the CPU many of these codes resemble
more general enterprise or ``cloud'' applications~\cite{CLOUDSUITE,GOODACHEP,GOODACYCLEACCT}.

Some of this is simply the legacy of code design during the long era
in which single sequential application performance scaled with Moore's
Law and the emphasis was on the use of a single, mostly homogeneous,
architecture.  When we began using commodity (ia32 and later x86-64)
processors, poor compiler support and the complexity of handling
multiple generations of processors in a grid environment also
discouraged the use of even the modest vectorization capabilities of
the processors. To first order, we programmed to a sequential,
least-common-denominator processor model to optimize the computing
price/performance ratio  and because we could expect that throughput would in
any case double every 1.5 years or so.

Our problems do, however, have many natural layers of parallelism (e.g. clusters,
hits, tracks) which could be exploited by explicitly parallel algorithms and data
structures. Similarly, through the appropriate design choices  it should be possible to
bring together sufficient floating point to exploit vectorization capabilities of
the processors much more than we do today. In some special cases such 
as simplified ``fast'' tracking in the trigger, likelihood fits during analysis, etc.
we should even be able to exploit the tremendous floating point capabilities 
of GPGPUs. Undoubtedly further possibilities will come out as code and algorithms 
are rethought in the light of different processor capabilities.

\section{Areas of Research and Development}

In this section, we describe what we will believe are the relevant elements of
an R\&D and upgrade program which is necessary to meet the challenges posed by the
new heterogeneous processor environment. A program to insure that our software will
be sufficiently scalable and efficient on these architectures can be seen as an
upgrade effort through the end of this decade, accompanying the phase-1 LHC
detector upgrades, and an R\&D effort towards the eventual software we will
need for the phase-2 upgrades and HL-LHC.  This research will also
guide future hardware purchases.

\subsection{General Investigations of new Architectures}

In order to best understand how the new architectures can benefit
HEP experiments, we need to understand their individual characteristics
and how they are intended to be used. These can differ greatly
across different types of accelerators. For instance, GPUs differ
from accelerators such as the Intel Xeon Phi in that to use GPUs
requires explicitly coding  in the single instruction, multiple
data (SIMD) programming paradigm. In this paradigm, the parallel
nature of tasks must be exposed by the programmer (rather than
automatically discovered by the compiler.)  In SIMD, each of the
hundreds of threads executing in parallel executes the same
instructions, but on different data.  
The Intel MIC architecture can be used in a more
conventional programming model where the compiler uses heuristics
to uncover possible gains from parallel tasks. Coding in this
platform is likely to be easier than GPU coding and require less
specialised knowledge; however, it is also likely that the possibilities
for gain when the parallelism inherent in the problems is exposed
by the programmer are greater for the appropriate class of problems.
While on paper the two architectures (MIC and GPU) have a similar
number of vector processors, achieving the full potential of these
resources will likely require very different approaches.

Similarly, to get the most out of these accelerators, we need to
understand what constraints the computer memory architecture puts
on the data structures we use for our data. For instance, in GPUs,
performance can be greatly increased if access to memory is coalesced,
\emph{i.e.,} if nearby threads access data in adjacent memory
locations. These sorts of optimizations require a rethink of how
algorithms are written and data is organized.

The topics that need to be investigated range from the types of
architectures to consider (ARM vs Intel x86 vs GPUs), the types of
programming tool sets and languages (\textsc{opencl, cuda, thrust,
 tbb, openmp}, Google \textsc{go}),
the models of parallelism (implicit, \emph{i.e.}, discovered by
automated tools, vs explicit, \emph{i.e.,} exposed by the programmer).
Much research has been done in the Computer Science (CS) community
to discover patterns and algorithms for parallel computation, and
another task is to understand what has been done in the CS community
and how these patterns can be mapped onto HEP problems.


\subsection{High Level Trigger, Offline Reconstruction and Analysis}
Many of the important algorithms used in the HLT and the offline 
event reconstruction, as well as some aspects of the data analysis, are such
that their cost in CPU time increases 
non-linearly with luminosity and in particular with the combinatoric effects
resulting from increases in the number of pile-up events. For this reason,
in the future we expect their relative importance to the overall cost to
increase, and thus we expect that significant effort will be necessary
here. Eventually many portions of the code will need development
to achieve the necessary scalability and efficiency. For example, two concrete
and important areas which will need to be addressed are:

\textbf{Tracking:} Charged particles traversing the CMS tracker
leave behind energy
deposits (hits), which are recorded by the electronics.  Track
reconstruction uses these hits to measure the parameters of the
charge particles, including momentum.  The current track reconstruction
in CMS is sequential in nature.  Initial tracks (seeds) are made
from 2-3 hits, usually from the innermost (pixel) detector.  The
track building traces each seed through the detector layers, searching
for hits at each layer, and updating the track parameters at each
layer.  The building stops when it fails to find hits on two layers
or it reaches the end of the detector.  After all seeds have been
traced, the tracks are checked for duplicates.  The remaining tracks
undergo a final fit to establish the best track parameters.  The
CPU time taken by seeding, track building, and track fitting is
roughly divided 25\%, 50\%, and 25\%, respectively.  As the track
reconstruction was responsible for about 50\% of the CPU time used
by the CMS reconstruction in 2012, it is a logical choice for parallelization
efforts.  There are several possible approaches for parallelizing
the track reconstruction.

The simplest solution is to parallelize the seeding, building, and
fitting steps individually.  For the seeding, the detector can be
divided into regions while for the track building (fitting), the
input seeds (tracks) can be divided into groups.  One downside to
this approach is the necessary synchronization after seeding and
building to check for duplicate tracks.  This approach is minimally
invasive and provides the necessary scaling for additional full
function cores.  However, each process will still need to perform
a variety of complicated instructions, limiting the applicability
of this approach to systems with multiple simple cores.

There are alternative track reconstruction algorithms which are
more inherently parallel and which can take advantage of many simple
cores and vector instructions.  Two examples of such approaches are
the Hough transform and cellular automata.  The Hough transform
works by applying a conformal transformation to all hits in the
detector such that all hits belonging to the same track cluster in
a well defined way.  Track finding amounts to locating clusters of
points rather than iteratively traversing detector layers.  The
conformal transformation and the cluster finding should be good
candidates for vectorization and parallelization with simple cores.
The main drawback to the Hough transform comes from the assumption
that tracks follow perfect helices, which is only true in a uniform
axial magnetic field with no material.  Therefore, allowances need
to be made for multiple scattering and energy loss from material
and the effect of a nonuniform magnetic field.  One possible
implementation of a cellular automata approach is to find all
combinations of three hits that are consistent with a track and
then combine the triplets together.  The triplet finder can be made
simple, allowing it to be vectorizable, especially when effects of
material are ignored.  Furthermore, by using appropriate data
structures, the triplet combination process can be made simple as
well and only depend on information in nearby memory locations.
This makes it suitable for a system with many simple cores, each
with dedicated memory.

The current track reconstruction performed during offline processing
of the data is similar to that performed by the high level trigger
(HLT) and the current hardware is also similar.  Thus, most changes
to the track reconstruction will provide benefits to both regimes.
However, at the trigger level the timing concerns are much more
critical.  Therefore, in the future, the hardware used by the HLT
(or earlier trigger) may become more specialized and be the first
to take advantage of the new architectures. Thus, the trigger may
become the logical testbed for new implementations of the track
reconstruction.

\textbf{Jet Clustering:} The energetic deposits of charged and neutral hadrons in the hadronic
calorimeter, as well as the deposits of electrons and photons in the
electromagnetic calorimeter and the deposits of charged hadrons
described above, (which are combined into a single ``particle flow
candidate'' at CMS) need to be clustered to obtain the complete
response. This is because the process inherently involves a ``shower''
of particles that spreads in both the lateral and radial directions,
called a ``jet''. In essence, this involves nearest-neighbor clustering
as described in detail in Ref.~\cite{FASTJET_TIMING}. Currently the
computational time is $O(N^2)$ or $O(N lnN)$, where $N$ is the input
number of separate detector elements encompassed by the shower (which
scales linearly with the number of particles per unit area, and hence
the luminosity).

At CMS, this computation is done very often by individuals accounted
in the 40\% of CPU usage from ``mixed user analysis applications'' as
described above. It is likely, therefore, that improvements observed
in jet clustering will primarily benefit this portion of the CPU
usage.

One remaining piece that may benefit dramatically from
parallelization, however, is the estimation of the area of these jets.
Currently, the procedure implemented at CMS is to randomly introduce a
large number of infinitesimally soft candidates called ``ghosts''
throughout the area considered. These are then globally added to the
jet clustering, and the locations of each are recorded. At the end,
the included candidates can be used to estimate the size of the jet
(with a resolution dependent upon the number per unit area of the
ghosts). Recent developments~\cite{FASTJET_MANUAL} of computing this
``ghosted area'' on a local basis rather than on a global basis could be
parallelized to reduce the computational time by factors of N(cores).

In addition, nontrivial extensions of the nearest-neighbor clustering
could be utilized to take advantage of parallelization. This requires
some algorithmic development to understand the potential gains and any
possible solutions.

\subsection{Parallelization of Geant4}

As described above, simulation with \textsc{geant4} is a very important fraction
of the total CPU used by CMS, as well as many other HEP experiments. It has,
and continues to be, the focus of optimization activities by a number of
groups, including CMS. As a toolkit used by many in the HEP community,
supported by a worldwide collaboration, there are some existing efforts in
a number of places, including CERN, SLAC, FNAL and others to understand
how to evolve \textsc{geant4}. Here CMS does not need to do the 
Geant4 development ourselves. We instead need to insure that we will
benefit maximally as new Geant4 versions are integrated with
the CMS Framework and are used with the CMS detector geometry, physics model
choices, etc. We should engage with, and provide feedback to, the teams 
working on Geant4 during the development period to guarantee that
outcome.

The currently ongoing efforts by the \textsc{geant4} team at SLAC to
integrate changes from
a thread-parallel prototype prepared by a team from Northeastern
University~\cite{GEANT4MT} are an important first step, as well as
investigations by the FNAL \textsc{geant4} team of the use of GPUs.
Taken together these two efforts represent one of the broadest and
potentially
highest impact R\&D programs into the newer architectures in the field.
This work on Geant4 will undoubtedly inform other efforts.

We note in addition another important potential gain from a scalable, multithreaded
\textsc{geant4} simulation which eventually supports a heterogeneous range of hardware.
As simulations based on \textsc{geant4} typically have rather modest input data
requirements (relative to data reconstruction, for example), and 
significant CPU use, they are perfect candidates for exploiting 
``opportunistic'' resources or even ``volunteer computing''. In opportunistic
computing, we use computing clusters owned by others and often designed
for other types of workflows. In volunteer computing, private individuals
donate spare computing cycles, typically on their desktop machines. In
both cases, the ability to use whatever processor hardware is available,
while simultaneously limiting as much as possible memory use and the 
amount of output data held resident on the host machine, will maximize
the potentially usable resources.

\subsection{Tools and Systems}

In the next years, CMS will need to both develop new code and re-engineer
existing code to perform in a scalable fashion on the new architectures.
This will require evolving changes to the core software framework and the
related code infrastructure which provides the processing and event data
model used by CMS. An implementation of the CMS Framework allowing
for parallel (threaded) execution of the existing Framework modules
on multiple x86-64 cores is being prepared for use for the LHC run
starting in 2015. In
the following years, we expect this to evolve significantly as we 
re-engineer the algorithms and data structures to bring out the more fine 
grained parallelism required for scalable and efficient use of the
new processors. Depending on the results of the investigations mentioned
above, additional technologies like OpenCL, CUDA and/or others still to
appear will need to be introduced and interfaced with the full processing framework. 
Given that on the time scale of HL-HLC it is likely that several orders of magnitude
increase in parallelism may be required, the tools available today will
surely not be the last word.

There will surely be significant evolution in compilers and associated tools,
and perhaps also on operating system support. In addition some number of
auxiliary support tools for
performance profiling, code analysis and debugging will be required. Given
the more complex nature of parallel programs and the sorts of specialized
architectures we may see, code profiling will be an important activity to
insure efficient use of the processors and find limits to scalability. General
purpose tools like IgProf \cite{IGPROF}, as well as more advanced and specialized 
profilers, will likely be needed more widely. Similarly run-time debugging
applications which will be much more complex than today's simple sequential
applications will need more advanced tools, which simplify the process will be critical.
In order to sure code quality and find potential faults early for newly
written code, we also expect that code analysis tools will become increasingly
important to our code development and integration process.

One important class of support tools is math libraries. During the
frequency scaling era for CPU's and in particular with the transition
to object oriented programming, general interest in the numerical
aspects of programming arguably waned somewhat. This is changing.
The combination the transition to x86-64 (which brought the transition
from x87 to SSE2 floating point), more dynamic open source
compiler development (gcc since version 4, LLVM) and the need to use
the new architectures has renewed the interest in numerical 
computing, vectorization, etc. One example of this is the VDT~\cite{VDT}
library which provides inlineable, vectorizable versions of
mathematical functions and allows tradeoffs between accuracy
and speed. We will need to investigate the use of this and other
such tools as well as the numerical aspects of our own code
to understand when and where they can be used.

Large, distributed collaborations like CMS require tools that can be used
in many environments, including large labs like CERN and FNAL, but also 
universities, grid computing centers, clouds and other places where 
CMS collaborators may do their day to day work. Thus we have a strong
preference for Open Source software, however the nature of these
tools is such that using them is often a participatory process. When we 
work at the edge of the technology envelope, at scales and in ways that
others do not, we often use such tools to their limits. We have a history of
making contributions to open source tools, both by contributing bug fixes
and new features, but also by testing ``at scale'' the very latest development
versions of these tools and systems. Continuing and perhaps expanding
our activities in this area will likely be necessary to insure that we have the
tools we need for our environment.

Eventually the software itself should be able to manage heterogeneity
at the level of an individual worker node, however both during
the initial introduction of new architectures and in the long run it
will still be necessary to manage the heterogeneity between
clusters of resources on the grid. Support for such heterogeneity
will be needed both at the level of the CMS workflow management
tools and in the grid software.

\subsection{Input and Output of data (I/O)}
Although we focus primarily on processor technology in this paper, we note
that I/O concerns are also relevant here in two different ways. First, simply
feeding sufficient input data to the individual processor units and insuring that
outputs created are collected from
them in an efficient, scalable way is likely to be a major challenge as the required
parallelism increases by orders of magnitude.
For example, today's sequential applications are typically designed to expect
explicit serialization points
for outputs, which will likely cause scalability problems as we move to
highly concurrent, parallel applications.

Second, in order to use vector units efficiently and to reduce stalls resulting from
accessing data not in the processor memory caches, there will be more
emphasis on data structure layout and data locality than in the (object oriented)
past. Simpler data structures such as structures of arrays (SoA) and arrays
of structures (AoS) will likely become more common and software algorithms
will need be adapted. 

Indeed these changes in data structures will sufficiently different that
simpler and more standard I/O layers might also be considered for the large scale
and performance intensive portions of our computing workflows.

\subsection{Collaborations}

The problems presented by the new heterogeneous architectures and
the resulting requirements on software are not unique to HEP. While
we will almost certainly have to rethink adapt our own algorithms,
data structures and software designs at some level, and the nature
of advanced research also tends to put us at the bleeding edge of
new technologies, it is unlikely that HEP alone will provide complete
solutions to these problems. We will certainly profit from collaborating
with others and in this section we emphasize several types of
collaborations which we expect will be important.

\textbf{Other experiments and scientific projects:} There are a
number of large scientific HEP, nuclear physics and astronomy
projects with a major stake in the evolution of computing in the
next decade or more. We of course most resemble Atlas at the LHC,
but also Alice/LHCb at the LHC, the FAIR experiments at GSI, Belle
II at KEK, IceCube at the South Pole and even cosmic frontier projects 
like LSST will require computing at scales such that they either must 
or have already confronted these problems (see, for example, \cite{ICECUBEGPU}).
When we identify opportunities for practical technical collaborations
with other scientific projects, we will pursue them.  In particular,
the Lattice QCD community in both Italy and the United States has experience
with GPGPU accelerators, and we have reached out to learn from their experience
and possible collaboration~\cite{APE}. Some initial production use of
GPU's in HEP trigger environments by Alice~\cite{ALICEGPU} is also
an example where experience can be shared.
In addition
periodic teleconference meetings like the CERN Concurrency Forum,
dedicated workshops and conferences where we can exchange information
related to the development of parallel software/algorithms, tools
and the exploration of new architectures will of course also continue
to be important.

\textbf{Computer Science:}  In general HEP has had a number of
collaborations with computer scientists in the area of middleware,
but collaboration on topics related to the software has been somewhat
rarer.  The evolution of processor technologies and the overwhelming
trend towards parallelism in the software is however also driving
relevant computer science research. HEP collaboration with computer
scientists would potentially be a mutually beneficial arrangement.
HEP benefits from direct access to the accumulated body of knowledge
and state of the art in computer science research into parallel
processing and algorithms.  The scale of HEP software and computing,
and its openness, also provide opportunities for computer scientists
to work with large codes and ``real world'' problems, but free of
proprietary restrictions of the commercial world.  This  can inform
their research and push it into new regimes in ways that would
otherwise be difficult.

We envision several major areas in which collaboration might be
useful:  the development of new or improved algorithms more appropriate
for these architectures, tools to compile for, debug, profile or
evolve codes to these architectures and general investigations into
how to build extremely parallel software/hardware systems. Undoubtedly
additional topics will arise as we begin such collaborations.

\textbf{Industry:} First and foremost, we expect collaborations
with the hardware vendors (Intel, AMD, NVIDIA and others) will be
very useful. Some of these collaborations exist, e.g. via CERN
Openlab and FNAL, plus some contacts typically between the vendors
and the research computing centers of collaborating universities.
Some of these amount to simple opportunities for the vendors to
pass on information about their latest products and plans.  In some
cases, more collaborative relationships have been developed where
the vendor representatives become involved in a ongoing discussion
and collaboration more aligned with our specific needs and use
cases. It is somewhat easier to develop such collaborations in the
context of a formal relationship like CERN's Openlab. Indeed, it
would be an extremely good idea to explicitly explore similar types
of partnerships between FNAL and industry, with CMS directly in the
collaborative loop.

In addition to the hardware vendors, we have on occasion developed
useful relationships with individuals working for companies more
focused on software, including Google, Redhat, Oracle and others.
This has been a bit ad-hoc, sometimes the connection has been through
someone who previously worked in HEP, other times through common
participation in various open source software projects. As with the
hardware vendors, we should explore possible partnerships and
collaborations, up to and including some arrangements like CERN's
Openlab between FNAL and the companies with significant CMS 
involvement.

\subsection{Education and Training}

The techniques, tools and technologies needed to write efficient
and scalable software for these new types of processor architectures
will be more sophisticated than what has typically been required of the
standard physicist collaborator to do software development in the
past. Even if some core ``professional'' software support effort exists
within the collaboration, as described earlier we rely heavily on
software development contributions from the collaboration at large.
To maximally leverage the significant experience of these physicist
collaborators on CMS, we expect that some amount of education and
training will be needed.  This will likely range from introductory
training in parallel programming, and all related technologies chosen
by the experiment, to more complete short courses or schools such
as the ESC\cite{ESC12} series of schools organized by the INFN in
recent years in Italy. Overall the minimum skills and knowledge to
contribute to the experiment's software will likely be greater than before.
But as the CMS Data Analysis Schools (CMSDAS) \cite{CMSDAS} have
shown, by organizing explicit education and training on relevant topics
for our collaborators, we can make a significant impact.

The training of graduate students and postdocs is an intrinsic part
of a research enterprise like CMS. Insuring that they have cutting edge 
computational skills and experience, regardless of whether they continue
in HEP or move on
to other fields or industry, is a valuable investment for the long term.
Simultaneously, the possibility of participating in discovery research,
using and pushing the envelope with advanced tools is also an
important aspect of what attracts the best and the brightest students
to our field.

\section{Defining the Research Program and Deployment Scenarios}

In the previous section we described a complete set of research
elements which we believe will be required in the coming years.
Subsequent documents from various groups will make detailed research
proposals covering one or more of the needed research elements.
As noted earlier, we expect that over the next 5 years or so the
resulting activities
should consist both of work aimed at concrete upgrades based on
the available technologies and of more general R\&D activities aimed 
at understanding how to use the evolving technologies which will
become available on the time scale of HL-LHC. The deployment of
the newer technologies as upgrades to the current software
and computing system will allow for both overall performance 
improvements in the medium term and gaining experience which will
inform the R\&D work towards the long term.

An important consideration in preparing detailed research plans is
the possible timing of such upgrades relative to the CMS and LHC running 
schedule over the coming years. The most important thing to
note is that it will not be possible to upgrade the entire worldwide
computing system used by CMS (offline and HLT) to any given new
hardware technology
at a single point in time. Any such upgrades will inevitably lead
to a heterogeneous system and how we manage that is one part of the
problem. In the rest of this section we 
outline the constraints on, and opportunities for, such deployments.


The schedule for such a deployment
needs to balance several factors. First, once we understand the
improvements that can be achieved, it will guide us as to when each scenario
is mature enough to be deployed. 
Secondly, the deployment needs to be timed in accordance
to the impact on the running experiment (for improvements to the HLT
or T0 farms critical for data-taking, say) and in accordance to the
availability of funding (to be able to replace entire farms with
homogeneous farms based on new architectures.)  Intermediate scenarios
that can be deployed parasitically will also be examined.

One can consider several possible time-lines for installation of new
computing resources, based on their impact on data-taking and on the
LHC run schedule.
\begin{itemize}
\item \textbf{`Any time' installations in offline/grid computer centers:} For offline resources not directly involved in the running of the experiment,
e.g. not involved in Prompt Reconstruction, it will be possible to
deploy new hardware at any time. Any kind of farm (small core, accelerator,
 etc) could be tested this way using upgraded software for simulation, re-reconstruction or analysis
workflows and eventually brought into production use. Opportunistic use
of farms with new technologies owned by others is a similar case. It will
however be difficult
 to test
 solutions that are intended for on-line (\emph{i.e.}, HLT) in such a
 scenario. 
\item \textbf{Short shutdowns/year-end technical stops:} Here it will be possible to install
 parasitic systems for HLT or make some types of major changes to the Prompt Reconstruction
and re-reconstruction workflows in the T0, T1. Parasitic deployments of
 upgrades to trigger system have been demonstrated in past
 experiments (e.g., CDF) and are planned for the Phase~1 CMS trigger
 upgrades. The method works as follows. A copy of the data is sent to
 a second parallel system (the upgrade system under examination)
 which processes the data as if it were making the actual trigger
 decision, and the results examined against expectation. All inputs
 to the new system, including timing, is by construction identical to
 the actual system. Results can be re-injected into the data stream
 for offline analysis. Once the performance has been demonstrated,
 the switch-over to the new system can proceed quickly. This
 parasitic approach allows optimal use of short downtimes and can
be critical for gaining confidence and experience with new technologies
in such environments.
\item \textbf{LS1.5 or other long shutdowns:} Full replacement of big systems
 like the HLT and the largest changes for data-taking critical systems
like Prompt Reconstruction (in the T0 and T1's) will only be possible
during the longer shutdowns.
 These long down times would allow for extensive testing
 of new systems and also enough time to revert to previously known
 good systems in the case of problems.
\end{itemize}
Staged deployment ensures no loss of data-taking efficiency. Parasitic
deployment for systems necessary for data-taking minimizes risks. We
envision that we can take advantage of shorter down-times, and would
not have to wait for a long shutdown to deploy upgrades.

Another topic to investigate is what kinds of farms we could
deploy. Replacing entire farms is a costly endeavor; can we make do
with heterogeneous farms to seamlessly move from the current
configuration (without accelerators and many-core) to a new
configuration? An example list of types of installations is below.
\begin{itemize}
\item Single farms with a homogeneous configuration. This would be
 easiest and cleanest to handle. For example, farms with ARM
 processors, or uniformly configured with accelerators such as GPU or
 MIC. Such a farm would require a large upfront investment into
 hardware but would be straightforward from the deployment point of
 view, and would not require software to manage mixed
 configurations. Every job launched on this farm could expect the
 same hardware configuration and to run the same code. This most
 closely mimics what we have today.
\item HLT Pre-processor: standalone pre-processor for HLT idea, main
 farm remains unchanged. Spy on data from FEDs; peel off data from
 appropriate FEDs; do processing to produce ``Level 1.5'' trigger
 data, such as tracks or primary vertices, and re-inject this data
 into the data stream for processing by the full HLT farm. If this is
 feasible and provides good speed-up this could be done parasitically
 and possibly quickly. Net result would be a speedup of average HLT
 processing time. Such a scenario does not require changes to rest of HLT
 farm. Not clear what the speed-up would be; needs to be studied.
\item Heterogeneous farms with fractions of instances with
 accelerator(s). Example use cases:
 \begin{itemize}
 \item HLT: designate certain types of triggers to go to these specific
   DAQ resources; e.g., those with muon triggers or other
   tracking-intensive triggers.  
 \item Generic offline farm: Requirements on jobs if they do or do not
   require accelerators; possibly have jobs that can work in either
   case but just run more slowly w/o the accelerator. Speedup for
   overall processing; could be rolled out gradually.
 \end{itemize}
\end{itemize}
This is a short list of ideas on farm types that show the need for a
research program to examine how to use the emerging many-core
technologies. In all instances, care must be taken to ensure that the
running experiment is not negatively affected, and that the cost
profile of the deployment of new resources is realistic.

\section{Conclusions and Outlook}
Progress in particle physics has long been driven not only by new
and more powerful accelerators and detector technologies, but also
by exploiting maximally the exponential increase in computing performance
available per unit cost achieved by industry over many decades. The
underlying technology trend, famously described by Gordon Moore, appears
to be continuing. In the last decade, however, limitations due to
overall power consumption have made it more difficult to translate
the raw technology gains into actual increases in computing performance.

In this document we have described the problem, both generally
in industry and specifically in CMS and in our own field. We have
described the needed elements of the broad R\&D and upgrade program
we believe is necessary to affront the problem, both in the next
years and for the longer term (HL-LHC). Establishing and executing
such a program will be
critical to the success of CMS and HEP in general in the coming
years.


\newpage
\bibliographystyle{unsrt}
\bibliography{references}

\end{document}